\documentclass[linenumbers]{aastex631}
\usepackage{amsmath}

\shorttitle{Novel Monte Carlo Moves}
\shortauthors{B. Militzer}

\graphicspath{{./}{figures/}}

\newcommand{\rr}{\ensuremath{\vec{r}}}

\begin{document}
\nolinenumbers

\title{Ensemble Monte Carlo Calculations with Five Novel Moves}

\author{Burkhard Militzer}
\affil{Department of Earth and Planetary Science, Department of Astronomy,\\ University of
California, Berkeley, CA, 94720, USA}

\begin{abstract}\nolinenumbers
  We introduce five novel types of Monte Carlo (MC) moves that
  brings the number of moves of ensemble MC calculations from three to
  eight. So far such calculations have relied on {\em affine
    invariant} stretch moves that were originally introduced by
  Christen (2007), {\em walk} moves by Goodman and Weare (2010) and
  {\em quadratic} moves by Militzer (2023). Ensemble MC methods have
  been very popular because they harness information about the fitness
  landscape from a population of walkers rather than relying on expert
  knowledge. Here we modified the affine method and employed a simplex
  of points to set the stretch direction. We adopt the simplex concept
  to quadratic moves. We also generalize quadratic moves to arbitrary
  order. Finally, we introduce {\em directed} moves that employ the
  values of the probability density while all other types of moves
  rely solely on the location of the walkers. We apply all algorithms
  to the Rosenbrock density in 2 and 20 dimensions and to the ring
  potential in 12 and 24 dimensions. We evaluate their efficiency by
  comparing error bars, autocorrelation time, travel time, and the
  level of cohesion that measures whether any walkers were left
  behind. Our code is open source. 
\end{abstract}

\section{Introduction} \label{sec:intro}

Since the seminal paper by \citet{Me53}, Monte Carlo (MC) techniques,
in particular Markov Chain Monte Carlo (MCMC) methods, have been
employed in many areas of science including physics, astronomy,
geoscience and chemistry
\citep{Kalos,MonteCarlo2006,landau2021guide}. These methods are
particularly effective when states in a high-dimensional space need to
be studied but only a small but nontrivial subset of the available
states are relevant. The distribution of $n$ gas molecules in a
three-dimensional room is one example to illustrate their
effectiveness. A MC simulation would only generate configurations in
the $3n$ dimensional space in which the molecules are more or less
evenly distributed in the room, as one would expect. Formally with
analytical statistical methods, one would be required to also
integrate over highly unlikely configuration, in which, e.g., many
molecules are concentrated in one corner of the room. Such states are
irrelevant because they only occupy a negligible measure of the
accessible configuration space. MC methods are very good in avoiding
unlikely configuration and in sampling only a representative subset of
the relevant configurations, which makes such MC method orders of
magnitude more efficient than exhaustive integration
methods. Consequently, MC methods offer no direct access to the
entropy nor to free energies but these quantities can be derived
indirectly via thermodynamic integration
\citep{Wijs1998,Wilson2010,Wilson2012a,Wahl2013,Gonzalez2014}.

In the field of statistical physics, MC methods have thus found
numerous applications to sample the states of classical and
quantum systems~\citep{binder2012monte,ceperley1995quantum,MP02,Driver2015} that
are weighted by the Boltzmann factor,
\begin{equation}
\pi(\rr) = \exp \left\{ - \frac{E(\rr)}{k_B T} \right\}\;\;.
\label{Boltz}
\end{equation}
$\pi$ is the unnormalized probability density of states, $\rr$, with
an energy, $E(\rr)$. The product of temperature, $T$, and Boltzmann
constant, $k_B$, controls the relative weights of high- and low-energy
states. Throughout this article, $\rr=\{r_1, \ldots, r_N\}$ represents
a vector in the $N$ dimensional space of states that is to be
sampled. We often refer to such vectors as {\it walkers}. $\rr_i$
refers to the $i$th walker in an ensemble while when we discuss the
specific sampling functions like the Rosenbrock density or the ring
potential, $r_m$ represents the $m$th component of a vector $\rr$. It
should be noted that a walker does not play the role of a particle in
physics-based MCMC calculations where typically $n$ particles move in
3D space, which leads to a $N=3n$ dimensional sampling space. In such
applications, one typically moves one or a handful of particles with
specialized types of moves that are particularly efficient for
specific interactions. The MCMC methods that we discuss in this
article are not focused on sampling particle coordinates but are more
general. They should for example allow one to construct ensembles of
models of Jupiter's interior for which a walker may represent the
masses and composition of certain
layers~\cite{Militzer_2024_23456}. So we do not assume that the $N$
dimensional sampling space can be divide into $n$ equivalent 3D
subspaces. When one moves a particle in physics-based simulations,
only its 3 coordinates change while when we move a walker here, all
its $N$ elements are affected.

Besides statistical physics, there is another broad area of
applications where MCMC are typically employed. In fields like
astrophysis, one often constructs models to match sets of telescope
observations or spacecraft measurements
\citep{hubbard-annrev-02,Bolton2017}, $y_i$, that come with a
statistical uncertainty, $\delta y_i$. MCMC calculations are then used
to map out the space of model parameters, $\rr$, that is compatible
with the observations by sampling probability
density~\citep{MilitzerSaturn2019,Movshovitz_2020,DiluteCore},
\begin{equation}
\pi(\rr) = \exp \left\{ -\chi^2/2 \right\} {\;\;\rm with\;\;} \chi^2 = \sum_{i=1}^N \left( \frac{y_i^{\rm model}(\rr) - y_i^{\rm obs.}}{\delta y_i} \right)^2
\label{xi2}
\end{equation}

The affine invariant MCMC method by \citet{Christen} and
\citet{Goodman} along with its practical implementation by
\citet{emcee} have been very popular to solve such kind of sampling
problems. Many applications came from the field of astrophysics. For
example, the affine sampling method has been employed to detect
stellar companions in radial velocity catalogues~\citep{PW2018}, to
study the relationship between dust disks and their host
stars~\citep{AR2013}, to examine the first observations of the Gemini
Planet Imager~\citep{MG2014}, to analyze photometry data of Kepler's
K2 phase~\citep{VJ2014}, to study the mass distribution in our Milky
Way galaxy~\citep{MP2017}, to identify satellites of the Magellanic
Clouds~\citep{KP2015}, to analyze gravitational-wave observations of a
binary neutron star merger~\citep{DF2018}, to constrain Hubble
constant with data of the cosmic microwave background~\citep{BV2016},
or to characterize the properties of M-dwarf stars~\citep{MF2015} to
name a few applications. On the other hand,
\citet{HuijserGoodmanBrewer2022} demonstrated that the affine (linear)
stretch moves exhibit undesirable properties when the Rosenbrock
density in more than 50 dimensions is sampled.

The MCMC in this article are not taylored towards any specific
application nor do they require knowledge of any derivitatives of the
sampling function $\pi(\rr)$ because they may very difficult to
compute for complex applications. This means however that the
algorithm has little information to distinguish favorable direction to
move into from unfavorable ones, and this is why one employs an
ensemble of walkers.

Ensemble MC methods propagates an entire ensemble of walkers rather
than moving just a single one. While this seems to imply more work,
such methods often win out because they harness information from the
distribution of walkers in the ensemble to propose favorable moves
that have an increased chance of being accepted without the need for a
detailed investigation of the local fitness landscape as the
traditional Metropolis-Hastings MC method requires. Many extensions of
the Metropolis-Hastings approach have been
advanced~\citep{AndrieuThoms}. For example, \citet{AdaptiveMH} use the
entire accumulated history along the Monte Carlo chain of states to
adjust the shape of the Gaussian proposal function.

Ensembles of walkers are employed in
various types of Monte Carlo methods that have been designed for specific
applications. In the fields of condensed matter physics and quantum
chemistry, ensembles of walkers are employed in {\em variational}
Monte Carlo (VMC) calculations~\citep{martin_reining_ceperley_2016}
that optimize certain wavefunction parameters with the goal of
minimizing the average energy or its variance~\citep{FM99}. Ensembles
are used to vectorize or parallelize the VMC calculations. They are
also employed generate the initial set of configurations for the
walkers in {\em diffusion} Monte Carlo (DMC) simulations. In DMC
calculations, one samples the groundstate wave function by combining
diffusive moves with birth and death processes. An ensemble of walkers
is needed to estimate the average local energy so that the birth and
death rates lead to a stable population size. Walkers with a low
energy are favored and thus more likely to be selected to spawn
additional walkers. Walkers in areas of high energy are likely to die
out.

The birth and death concepts in DMC have a number of features in
common with genetic algorithms that employ a population of individuals
(similar to an ensemble of walkers). The best individuals are selected
and modified with a variety of approaches to generate the next
generation of individuals~\citep{schwefel, Militzer1998}. The
population is needed to establish a fitness scale that enables one to
make informed decisions which individuals should be selected for
procreation. This scale will change over time as the population
migrates towards for favorable regions in the parameter space. This
also occurs in DMC calculations as the walker population migrates
towards regions of low energy, the average energy in the population
stabilizes, and the local energy approaches the ground state energy of
the system.

Ensembles of individuals/walkers are not only employed in genetic
algorithm but are used in many different stochastic optimization
techniques. These methods have primarily been designed for the goal of
finding the best state in a complex fitness landscape, or a state that
is very close to it, rather than sampling a well-defined statistical
distribution function as Monte Carlo method do. Therefore these
optimization are much more flexible than Monte Carlo algorithms that
typically need to satisfy the detailed balance relation for every
move~\citep{Kalos}.

The particle swarm optimization
method~\citep{KennedyEberhart1997,EberhartShi2001} employs an ensemble
(or swarm) of walkers and successively updates their locations
according to a set of velocities. The velocities are updated
stochastically using an inertial term and drift terms
favor migration towards the best individual in the population and/or
towards the global best ever generated.

In general, efficient Monte Carlo methods are required to have two
properties. 1) They need to migrate efficiently in parameter space
towards the most favorable region. The migration (or convergence) rate
is typically measured in Monte Carlo time (or steps). 2) Once the
favorable region has been reached and average properties among walkers
have stabilized, the Monte Carlo method needs to efficient sample the
relevant parameter space. The efficiency of the algorithm is typically
measured in terms of the autocorrelation time or the size of the error
bars. 

\citet{Militzer_QMC_2023} recently introduced quadratic MC (QMC) moves and
demonstrated that they are more efficient in sampling challenging,
curved fitness landscapes than linear stretch
moves. \citet{Militzer_QMC_2023} also modified the {\em walk} move method by
\citet{Goodman} by adding a scale factor, which enables one to control
the step size of the moves and thus enable one to substantially
increase the MC efficiency. All these method propagate an ensemble of
walkers and harness information from their location about the fitness
landscape rather relying on expert knowledge to chose the promising
directions for future MC moves. The goal of this article is to add
five types of novel MC moves to the portfolio of available ensemble MC
moves because so far there are only three: affine stretch moves that
were first introduced by \citet{Christen}, walk moves by
\citet{Goodman}, and the quadratic moves by \citet{Militzer_QMC_2023}. Here we also
design a method to compare the efficiency of all these moves and then
apply it by conducting simulations for the Rosenbrock density in 2 and
20 dimensions and the ring potential in 12 and 24 dimensions. In
general, we do not know what types of moves will prove to be most
effective when MC methods are applied to a broad range of problems, so
having access to a portfolio of MC moves may prove useful in saving
computer time and thus energy.

\section{Methods} \label{sec:methods}

\begin{figure}[ht!]
\gridline{\fig{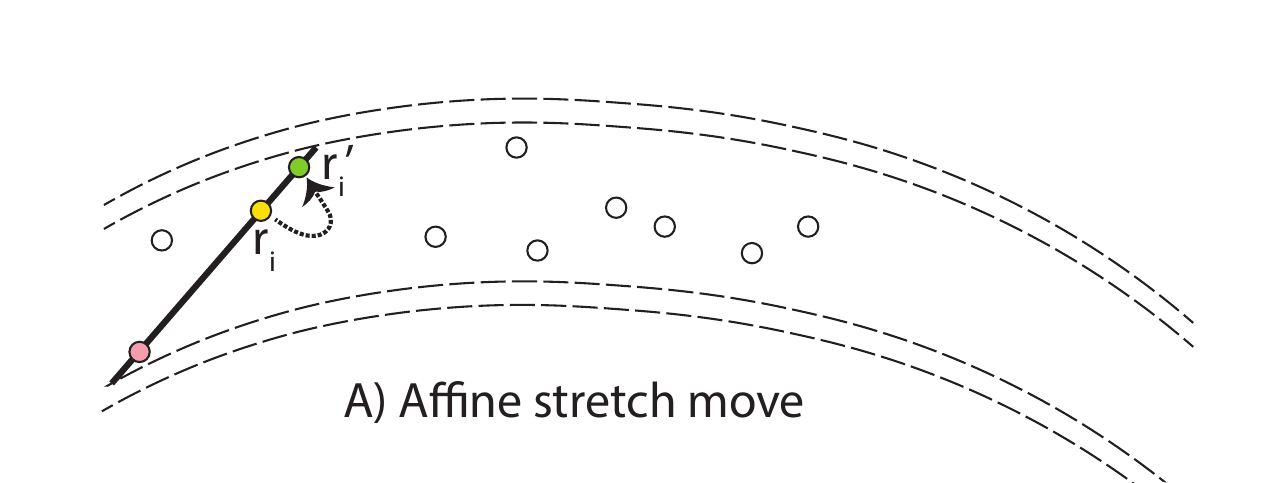}{0.45\textwidth}{}
          \fig{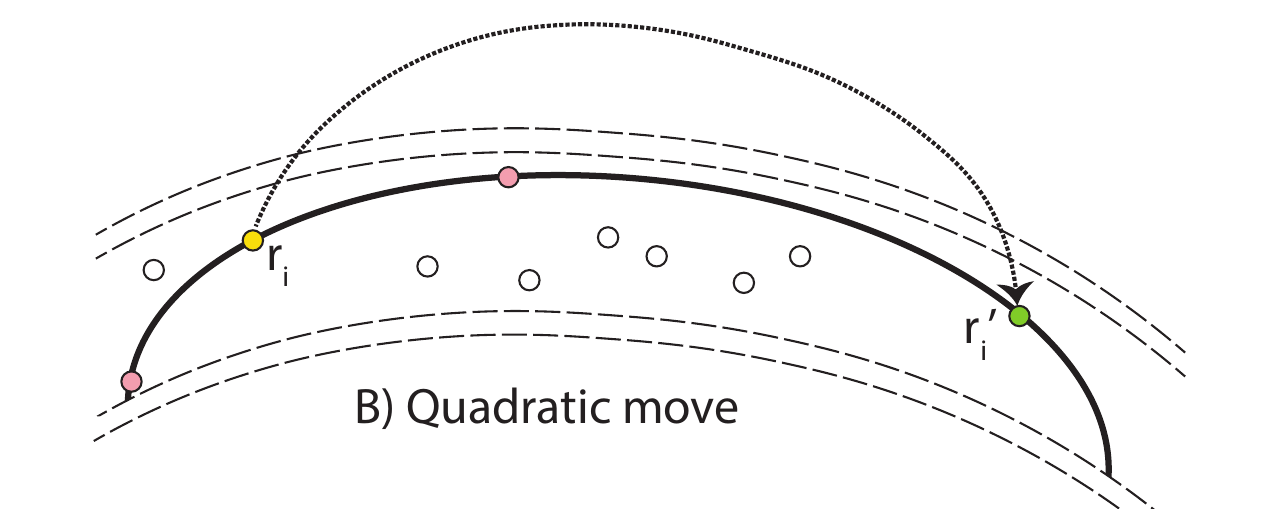}{0.45\textwidth}{}
          }
\gridline{\fig{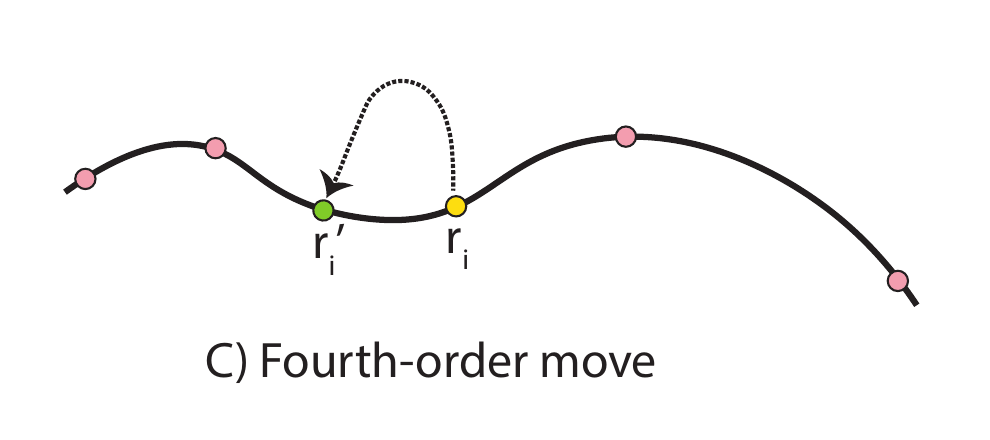}{0.38\textwidth}{}
          \fig{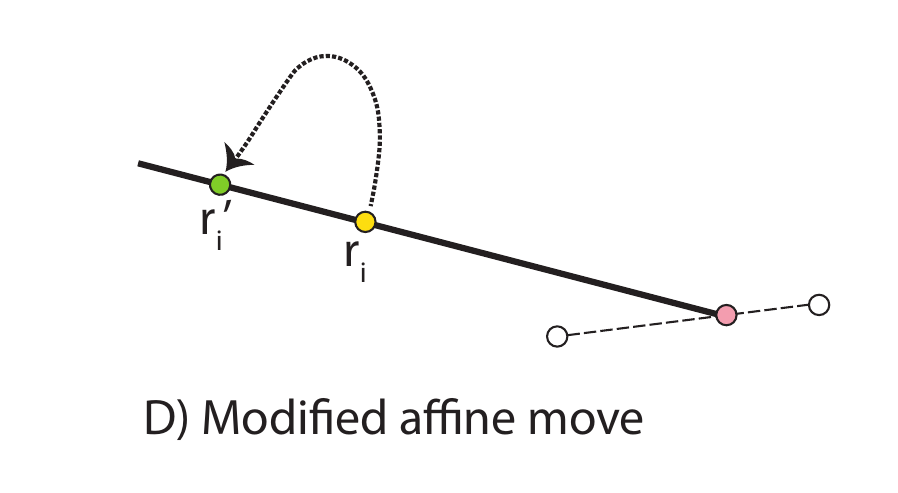}{0.38\textwidth}{}
          }
\gridline{\fig{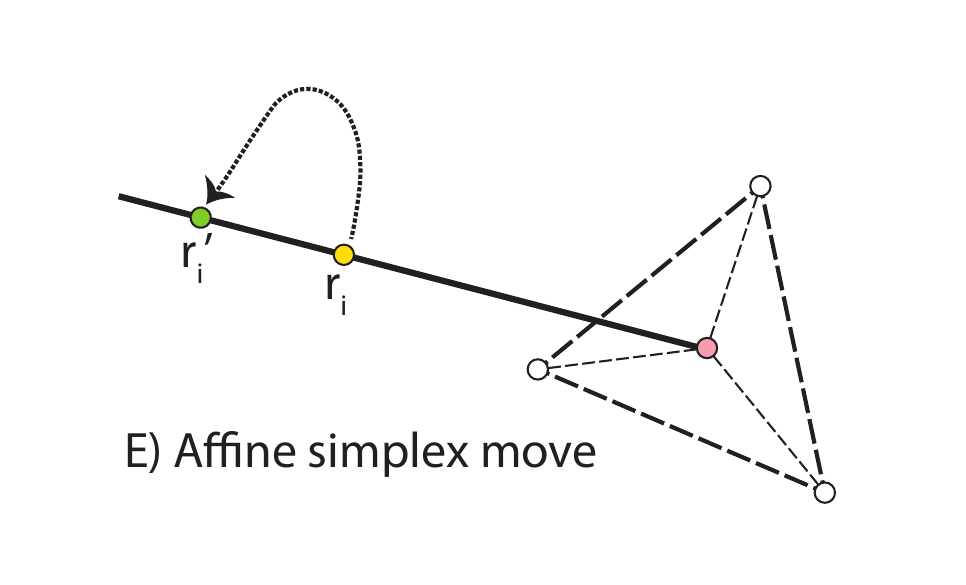}{0.38\textwidth}{}
          \fig{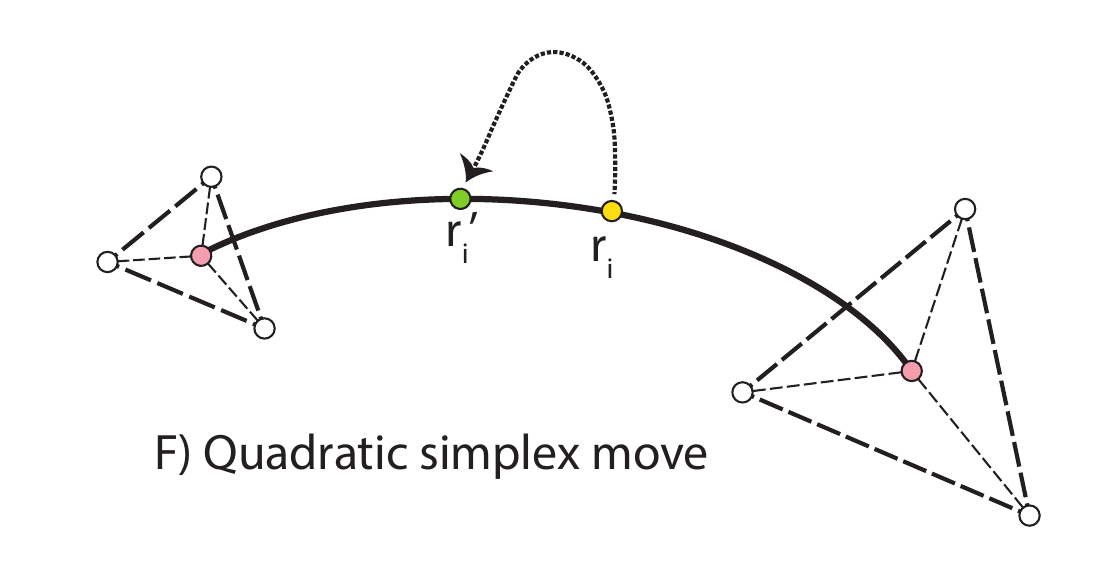}{0.42\textwidth}{}
          }
\gridline{\fig{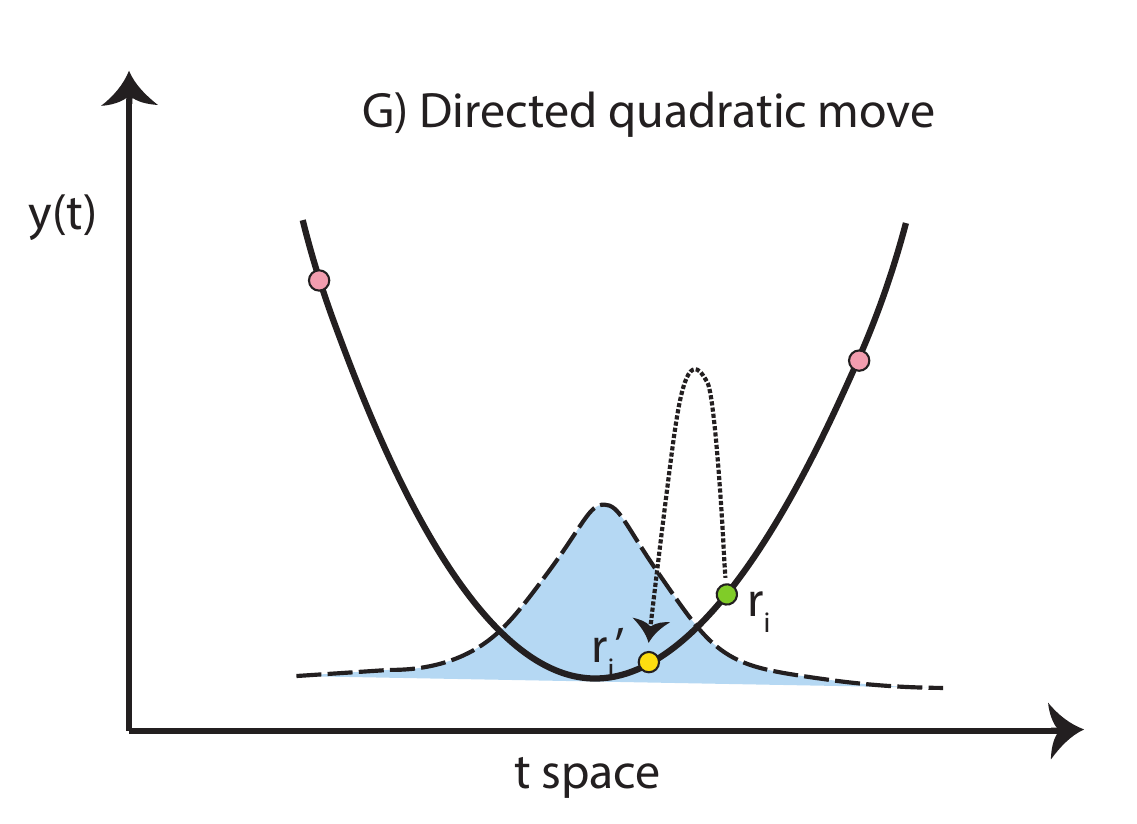}{0.45\textwidth}{}
          }
\caption{\label{fig:moves} Illustration of different types of MC moves. The affine moves from \citet{Christen} and quadratic moves from \citet{Militzer_QMC_2023} are shown in panels A and B. The remaining panels illustrate the novel moves that we designed for this article. Panel C shows a fourth-order move that employes four guide points while quadratic moves use only two. In panel D, we depict a modified affine move, for which we interpolate between two walkers to obtain the guide point that defines the stretch direction. In panels E and F, we generalize this concept by deriving the guide points by taking the center of mass from a group (or simplex) of walkers. Finally in panel G, we illustrate our {\it directed} moves that employed the energy, $y$, to fit a Gaussian function in order to sample favorable parameters regions more often. Conversely, all other moves only employ the location of the walkers but no information about their energy. (In this article, we also test and apply {\it walk} moves from \citet{Goodman} and {\it modified walk} moves from \citet{Militzer_QMC_2023} but we do not illustrate them here.)
}
\end{figure}

\subsection{Review of Affine, Quadratic, and Walk Moves}

To move walker $i$ from location $\rr_i$ to $\rr_i'$, \citet{Goodman}
employed the affine invariant moves that were originally introduced by
\citet{Christen}. From the ensemble, one selects one additonal walker,
$j$, to set the stretch direction,
\begin{equation}
                 \rr_i' = \lambda (\rr_i - \rr_j) + \rr_j
\label{affine}
\end{equation}
To make such moves reversible, the stretch factor, $\lambda$, must be
sampled from the interval $\left[\frac{1}{a},a\right]$ where $a>1$ is
constant parameter that we controls the step size. To sample the
stretch factor, $\lambda$, one has a bit of a choice. \citet{Goodman}
followed \citet{Christen} when they adopted a function,
$P_S(\lambda)$, that satisfies,
\begin{eqnarray}
P_S(\lambda) \;&=&\; \frac{1}{\lambda} \; P_S(\frac{1}{\lambda})\\
P_S(\lambda) \;&\propto&\; \frac{1}{\sqrt{\lambda}} \; \rm{if} \; \lambda \in \left[\frac{1}{a},a\right] \;\;.
\label{T1}
\end{eqnarray}
The acceptance probability is given by
\begin{equation}
  A(\rr_i \to \rr_i') = \min \left[ 1 , \frac{\pi( \rr_i')} {\pi( \rr_i)} \lambda^\alpha  \right]\;,
\label{AAffine}
\end{equation}
where the factor $\lambda^\alpha$ with $\alpha=N-1$ emerges because
one performs a one-dimensional move to sample points in a $N$-dimensions
space (see derivations in \citet{Christen} and \citet{Militzer_QMC_2023}).

In Fig.~\ref{fig:moves}, we illustrate the affine as well as the
quadratic moves. For a quadratic move, \citet{Militzer_QMC_2023} selects two walkers
$j$ and $k$ from the ensemble to move walker $i$ from $\rr_i$ to
$\rr_i'$.
\begin{equation}
\rr_i' = w_i \rr_i + w_j \rr_j + w_k \rr_k
\label{QMCmove}
\end{equation}
The interpolation weights $w$ are chosen from,
\begin{eqnarray}
w_i &=& L_2(t'_i \,;\, t_i, t_j, t_k) ,\\
w_j &=& L_2(t'_i \,;\, t_j, t_k, t_i) ,\\
w_k &=& L_2(t'_i \,;\, t_k, t_i, t_j) ,\\
L_2(x \,;\, x_0,x_1,x_2) &\equiv& \frac{x-x_1}{x_0-x_1} \frac{x-x_2}{x_0-x_2} \quad,
\end{eqnarray}
where $L_2$ is the typical Lagrange weighting function that guarantees
a proper quadratic interpolation so that $\rr_i'=\rr_i$ if $t_i'=t_i$;
$\rr_i'=\rr_j$ if $t_i'=t_j$; and $\rr_i'=\rr_k$ if $t_i'=t_k$. One
sets $t_j=-1$ and $t_k=+1$ to introduce a scale into the parameter
space, $t$.

To satisfy the detailed balance
condition, $T(\rr_i \to \rr_i') = T(\rr_i' \to \rr_i)$, it is key that we sample the parameters
$t_i$ and $t_i'$ from the same distribution $\mathcal{P_S}(t)$. The acceptance
probability then becomes,
\begin{equation}
  A(\rr_i \to \rr_i') = \min \left[ 1 , \frac{\pi( \rr_i')} {\pi( \rr_i)} \, \left|w_i\right|^N \right]\;.
\label{AQMC}
\end{equation}
Again a factor of $\left|w_i\right|^N$ is needed because we sample the
one-dimensional $t$ space but then switch to the $N$-dimensional
parameter space, $\rr$. In appendix of \citet{Militzer_QMC_2023}, this factor is
derived rigorously from the generalized detailed balance equation by
\citet{GreenMira2001}.

\citet{Goodman} also introduced {\it walk} moves. To move walker $k$ from $\rr_k$ to $\rr_k' = \rr_k + W$, one
chooses at random a subset of walkers, $S$. The subset size, $N_S$, is a free parameter that one needs to
choose within $2 \le N_S < N_W$ where $N_W$ is the total number of walkers. 
\citet{Militzer_QMC_2023} followed \citet{Goodman} in computing the average location all walkers in the subset,
\begin{equation}
\left< \rr \right> = \frac{1}{N_S} \sum_{j \in S} \rr_j \quad.
\end{equation}
but then modifoed their formula for computing the step size, $W$, by introduding a scaling factor $a$:
\begin{equation}
W = a \sum_{j \in S} Z_j \left( \rr_j - \left< \rr \right> \right) \quad.
\label{eq:W}
\end{equation}
$Z_j$ are univariate standard normal random numbers. By setting $a=1$,
one obtains the original {\it walk} moves, for which the covariance of the
step size, $W$, is the same as the covariance of subset $S$.
The introdcution of the scaling parameter, $a$, enabled
\citet{Militzer_QMC_2023} to make smaller (or larger) steps in
situations where the covariance of the instantaneous walker
distribution is a not an optimal representation of local structure of
the sampling function. It was demonstrated that the scaling factor $a$
significantly improves the sampling efficiency of the Rosenbrock
function and for the ring potential in high dimensions.

\subsection{From Quadratic to Order-N  Moves}

For a quadratic MC move one selects two guide points to move walker $i$
from $\rr_i$ to the new location $\rr'_i$. The new location is generated
by quadratic interpolation in parameter space $t$. Here we now
generalize this concept to arbitrary order, $N_O$. At random, we first
select $N_O$ guide points from the ensemble of walkers. Together with
the moving walker, we now have $N_O+1$ points to perform a Lagrange
interpolation of order $N_O$ to derive a new location for the moving
walker, $\rr_i'$,
\begin{equation}
\rr_i' = w_i \rr_i + \sum_{j=1}^{N_O} w_j \rr_j \quad.
\label{orderN}
\end{equation}
The interpolation weights, $w$, are given by
\begin{equation}
w_i = L_{N_O}(t',\vec{t},0) \;\;\;\;\; {\rm and} \;\;\;\;\; w_j = L_{N_O}(t',\vec{t},j)  \quad,
\end{equation}
where $\vec{t}$ is an $N_O+1$ dimensional vector
$\vec{t}=\{t_0,t_1 \ldots t_N\}$ that begins with index 0. $L_{N_O}$ are the
Lagrange polynomials,
\begin{equation}
L_{N_O}(x,\vec{t},s) = \prod_{i=1}^{N_O} \frac{x-t_{(i+s) \% (N_O+1)}}{t_i - t_{(i+s) \% (N_O+1)}} \quad,
\end{equation}
where the modular division in $t_{(i+s) \% (N_O+1)}$ guarantees that
index $i+s$ referes to vector element $t_{i+s}$ if $i+s<N_O+1$ and to
element $t_{i+s-(N_O+1)}$ if $i+s$ is larger. This leads to the usual
Lagrange interpolation, $\rr(t')$, that goes through all $N_O+1$
points so that $\rr_i'=\rr_i$ if $t'=t_0$ and $\rr_i'=\rr_j$ if
$t'=t_{j}$. For the quadratic moves in \citet{Militzer_QMC_2023}, we set
$t_{j=1}=-1$ and $t_{j=2}=+1$ to introduce a scale into the parameter
space $t$. The $t$ arguments for the original and new locations, $t_0$
and $t'$, are both sampled at random. 

\citet{Militzer_QMC_2023} proposed two alternative methods and sampled
them either a uniform distribution from $[-a,+a]$ or from a Gaussian
function with standard deviation $\sigma=a$ that is centered at
$t=0$. For the acceptance probabilities, we use again
Eq.~\ref{AQMC}. For both sampling functions, the parameter $a$
controls the average size of the MC steps in $t$ and $\rr$
spaces. While the standard deviation of resulting $t$ values is equal
to $a$ for the Gaussian $t$ sampling method, it is equal to only
$a/\sqrt{3}$ for the linear $t$ sampling. So the favorable ranges of
$a$ tend to be a little larger for linear $t$ sampling than for the
Gaussian method when the best settings of $a$ for both methods are
compared for the same application, as we will later see.

When we extend the quadratic moves to higher orders here, we still
sample $t_{j=0}$ and $t'$ as before but we need to specify the
arguments, $t_{j>0}$, for the remaining interpolation points. For a
third order interpolation, a natural choice would be $t_1=-1$,
$t_2=0$, and $t_3=+1$. Following this concept, we distribute the $t$
arguments uniformly,
\begin{equation}
t_{j>0} = 2\frac{j-1}{N_O-1} - 1
\end{equation}
so that $t_{j=1}=-1$ and $t_{j=N_O}=+1$. We study how this MC method
performs when we choose the order $N_O=\{3,4,6,10\}$ and combined it
with linear and Gaussian $t$ sampling and different values of
$a=\{0.1,0.3,0.5,1.0,1.2,1.5,2.0,3.0\}$. While we see a dependence on
$a$ when we compare the performance of this method, the performance of
the linear $t$ sampling method with the best choice $a$ was always
comparable to that of the Gaussian $t$ sampling method with its best
$a$. So for simplicity we combine the results with linear and Gaussian
$t$ sampling into one dataset when we later compare the performance
this method for different orders, $N_O$.

\subsection{Directed Quadratic  Moves}

None of the moves in \citet{Goodman} or \citet{Militzer_QMC_2023} nor any other
moves in this article make use of the probability values of the other
walkers, $\pi(\rr_j)$, when a new location, $\rr_i'$, for walker $i$
is proposed. Currently only their locations, $\rr_j$, are
employed. Some valuable information might be harnessed by utilizing that
a particular walker, $k$, resides in an unlikely location with a
probability value, $\pi(\rr_k)$, that much lower than that of the
others. (One may recall that the {\it regula falsi} method is more
efficient in finding the roots of a function than the bisection method
because it makes use of the function value while the bisection method
only relies on the sign of the values.)

Here we go back the quadratic moves that relies on the walker,
$\rr(t_0)$, and two guide points $\rr(t_1=-1)$ and $\rr(t_2=+1)$. We
assume their probabilities, $\pi(\rr(t'))$, came from an energy
function, $E(\rr(t')) = -k_BT*\log(\pi(\rr(t')))$. The factor $k_BT$
may be set to 1. We further assume that we can approximate the energy
function by the quadratic function, $E(t') = At'^2 + Bt' + C$, whose
coefficients are chosen so that it interpolates the three known
points, $E(t_0)$, $E(t_1)$, and $E(t_2)$. If $A$ is positive, it has a
minimum at $t_{\rm min} = -B/2A$ and we can sample $t'$ from a Gaussian function,
\begin{equation}
P_G(t') = \sqrt{\frac{A}{\pi k_BT}} \exp\left\{ -\frac{A}{k_BT}(t'-t_{\rm min})^2\right\}
\end{equation}
with the standard deviation, $\sigma = \left( 2A/k_BT \right)^{-1/2}$.

For the reverse move, one first needs to consider the probability of
sampling $t'$ from the distribution, $P_S(t')$. To sample $t$, one
constructs a different quadratic function, $P'_G(t)$, that
interpolates $E(t')$, $E(t_1)$, and $E(t_2)$. This introduces an
additional factor into the acceptance ratio,
\begin{equation}
  A(\rr_i \to \rr_i') = \min \left[ 1 , \frac{P_S(t')}{P_S(t)} \, \frac{P'_G(t)}{P_G(t')} \, \frac{\pi( \rr_i')} {\pi( \rr_i)} \, \left|w_i\right|^N \right]\;.
\label{DMC}
\end{equation}
The approach has one caveat because the interpolation coefficient,
$A$, may occasionally become negative during the forward or reverse
move, in which case the distribution $P_G(t')$ cannot be properly
normalized. When this happens if first try to reorder the elements of
the vector $\vec{t}$.  If this is not successful we overwrite $A$ with
$|A|$, which leads to a stable MC algorithm. We tested it with with
linear and Gaussian sampling, $P_S(t)$ and a series of $a$ values,
$a=\{0.1,0.3,0.5,1.0,1.2,1.5,2.0,3.0\}$.

\subsection{Modified Affine and Simplex  Moves}

Here we introduce three extensions of the affine and quadratic
moves. First for our {\em modified affine moves}, we linearly
interpolate between walkers $j$ and $k$ to pick a direction for the
stretch move of walker, $i$,
\begin{equation}
    \rr_i' = \lambda \rr_i + (1-\lambda) \rr_* {\;\; \rm with \;\;} \rr_* = c\rr_j + (1-c)\rr_k \;\;,
\label{modAff}
\end{equation}
where $c$ is a random number chosen uniformly between 0 and 1. In all
other respects, the move proceeds like the original affine
move. Illustrations of this and the two following moves are given in
Fig.~\ref{fig:moves}.

For our {\em affine simplex moves}, we set the direction of the
stretch move in yet a different way. As guide points, we select $N_G$
walkers from the ensemble, compute their center of mass, and insert it
as $\rr_*$ into Eq.~\ref{modAff}.
Finally for our {\em quadratic simplex move}, we select two separate
sets of $N_G$ walkers at random and compute their respective centers
of mass before inserting them as of $\rr_j$ and $\rr_j$ into
Eq.~\ref{QMCmove}. The move then proceeds like our quadratic moves.

\section{Applications} \label{sec:applications}

\subsection{Ring Potential} 

We test all our MC methods by applying them to two test problems: the ring
potential and Rosenbrock density. The ring potential is defined as,
\begin{equation}
V(\rr) = (2m)^{2m} \left[ (\rho-R)^{2m} + \sum_{i=3}^{N} r_i^{2m} \right] - C r_1\;\;,
\label{eq:pot}
\end{equation}
where $\rr=\{r_1, \ldots, r_N\}$ is a vector in the $N \ge 2$
dimensions. $\rho = \sqrt{r_1^2 + r_2^2}$ is the distance from the
origin in the $r_1$-$r_2$ plane. The first term ensures that the
potential is small along a ring of constant radius, $R=1$. The second
term keeps the magnitude of all remaining parameters,
$r_{3 \ldots N}$, small. Increasing the positive integer, $m$, makes
the walls of the potential around the ring steeper. For this article,
we performed calculations with $N=12$ and 24 dimensions while keeping
$m=6$ fixed. The last term breaks axial symmetry. We set $C$ to small
value of 0.01 so that the potential minimum is approximately located
at point $\vec{A}=(+R,0, \ldots)$ while the potential is raised at
opposing point $\vec{B}=(-R,0, \ldots)$. Figure~3 of \citet{Militzer_QMC_2023} shows
an illustration of this effect. The prefactor of the first term in
Eq.~\ref{eq:pot} is introduced so that the location of potential
minimum does not shift much with increasing $m$.

We employ the Boltzmann factor, Eq.~\ref{Boltz}, to convert the energy
of the ring potential into a probability function that we can
sample. We set $k_BT=10^{-4}$ so that the equilibrium distribution is
reasonably well confined around point $\vec{A}$. So when we compute
the autocorrelation time and the blocked error bar \citep{AT87} of the
potential energy, we initialized the ensemble of walkers around this
point. When we want to determine how long it takes the ensemble of
walkers to travel around the ring, we initialize the ensemble near the
high-energy point, $\vec{B}$. We define the travel time to be the
number of MC moves that are required for ensemble average of
$\left<r_1\right>$ to change sign from negative to positive. Based on
such a travel time analysis, \citet{Militzer_QMC_2023} concluded that employing
between $2N+1$ and $3N+1$ walkers was a reasonable ensemble size while
it took larger ensembles a disproportionally long time to travel
around the ring.

The ring potential also allows to define and study the {\em cohesion}
among an ensemble of walkers, which is relevant because all MC methods
in this article perform local moves that rely on information of
walkers in the ensemble. This means that if the walkers have split up
in a highly dimensional and fragmented fitness landscape, or it were
initialized that way, it will be challenging to reunited the walkers
later. This will likely to lead to a degradation of performance
because a distant walker cannot provide useful guidance for a local
move.
Therefore, we consider cohesion among the walkers to be a
favorable trait of an ensemble MC algorithm.

\begin{figure}[ht!]
\plotone{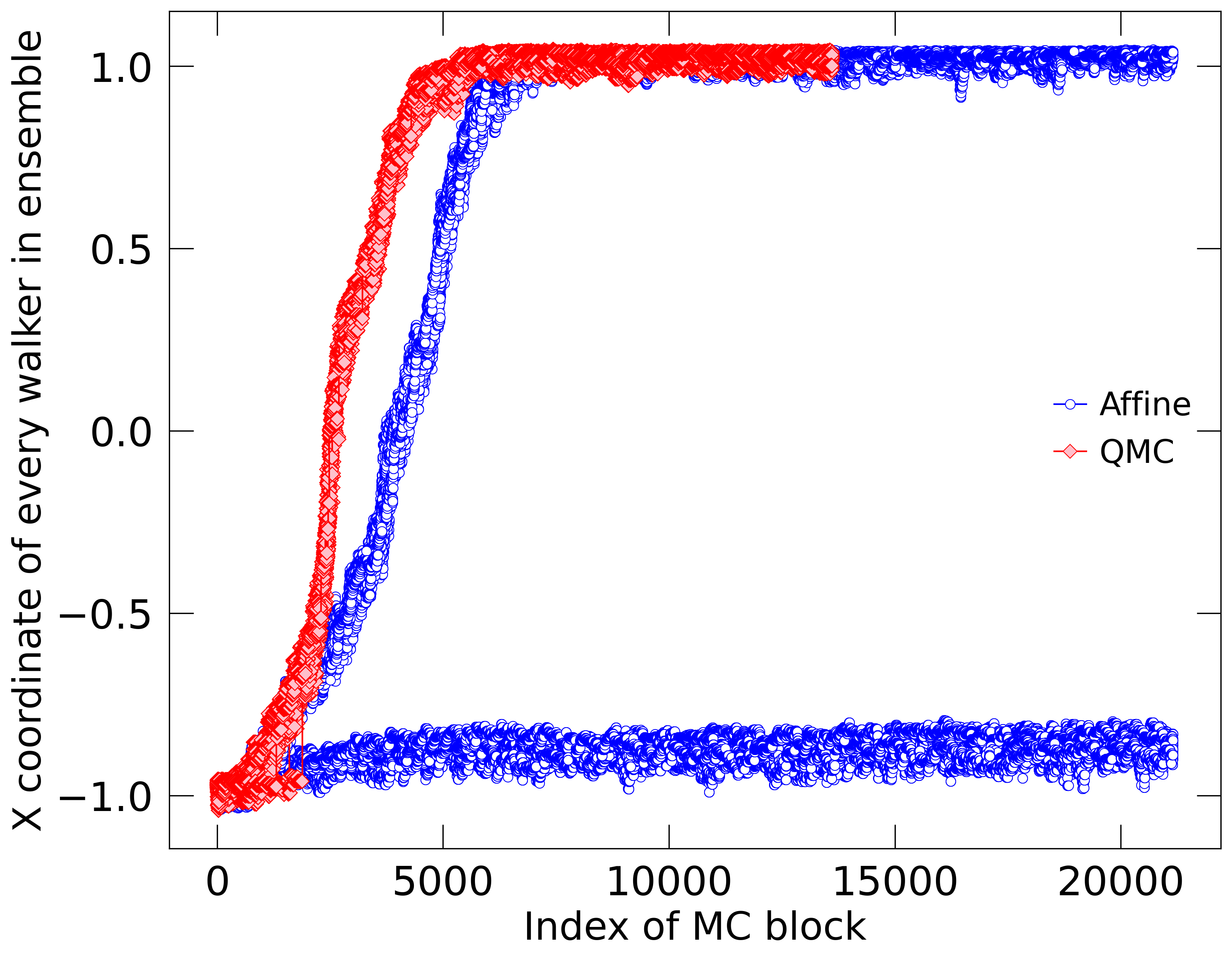}
\caption{To illustrate different degrees of cohesion, we plot the $r_1$ coordinate of all walkers at the end of every block. To compare the affine and QMC methods, we initialized two ensembles of $N_W=24$ walkers near the high energy state of $r_1=-1$ and monitored how long it took the ensemble to travel to the prefereed low-energy state at $r_1=+1$ and whether the ensemble remained together. The QMC ensemble traveled faster and walkers kept together. Conversely, the affine ensemble traveled a bit more slowly and split up so that only 17 of the 24 walkers arrived at the low-energy state.  \label{fig:cohesion}}
\end{figure}
 
For the ring potential, we define cohesion in the following, simple
way. As for the travel time calculation, we initialize the ensemble
near the high-energy state of point $\vec{B}$ and then monitor whether
all walkers find to their way to the low-energy state near point
$\vec{A}$. Cohesion is defined by the fracton of walkers with
$r_1 \ge 0$ at the end of the MC calculation. This average could in
principle depend on the duration of the MC calculations but
Fig.~\ref{fig:cohesion} provides an illustration why this dependence
is weak. Conservatively, we end our MC calculations after they
completed enough moves equal to twice the ring travel time.

To demonstrate why it is informative to compare the walker cohesion
between different MC algorithms, we plot the $r_1$ coordinates of all
walkers at the end of every block in Fig.~\ref{fig:cohesion}. For the
ring potential in $N=12$ dimensions, we initialized two ensembles of
with $N_W=24$ walkers near $r_1=-1$ and propagated them with our QMC
and with the affine methods. The QMC ensemble remained together and
traveled more efficiently towards the low-energy near $r_1=+1$. The
affine ensemble traveled more slowly and split up so that only 17 out
of 24 walkers reached the low-energy state. A careful inspection of
the first 2000 blocks reveals that even some walkers in the QMC
ensemble fell behind but eventually they all caught up. Still one may
not expect the QMC ensemble to show perfect cohesion in all
circumstances especially if sampling parameters are chose
poorly. While Fig.~\ref{fig:cohesion} shows just one calculation, we
always average the cohesion and the travel time over 1000 independent
MC calculations when we compare prediction from different MC moves and
parameter settings in this article.

\subsection{Rosenbrock Density} 

\citet{Goodman} tested their methods by sampling the 2d Rosenbrock density,
\begin{equation}
\pi_{2d}(r_1,r_2) = \exp \left[ \mathcal{R}(r_1,r_2) \right] {\;\; \rm with \;\;} 
\mathcal{R}(r_1,r_2) = - \frac{A \left( r_2-r_1^2 \right) + (1-r_1)^2}{B}\quad,
\label{Rosen2d}
\end{equation}
which carves a narrow curved channel into the $(r_1,r_2)$ landscape.
$B$ effectively plays the role of temperature while $A$ controls the
width of the channel. Here we set $A=100$ and $B=5$ to be consistent
with \citet{Goodman}. There are three different ways one can
generalize the Rosenbrock density to higher dimensions. First one can simply
treat it as $N/2$ independent 2d Rosenbrock problems by setting,
\begin{equation}
\pi_{\rm simple}(\rr) = \exp \left[ \sum_{i=1}^{N/2} \mathcal{R}(r_{2i-1},r_{2i}) \right] \;\;.
\label{RosenSimple}
\end{equation}
Alternatively one can ``connect'' the individual coordinates by defining,
\begin{equation}
\pi_{\rm connected}(\rr) = \exp \left[ \sum_{i=1}^{N-1} \mathcal{R}(r_{i},r_{i+1}) \right] \;\;.
\end{equation}
Finally one can make the problem periodic by also connecting the coordinates $r_1$ and $r_{N}$,
\begin{equation}
\pi_{\rm periodic}(\rr) = \exp \left[ \sum_{i=1}^{N} \mathcal{R}(r_{i},r_{1+i \% N}) \right] \;\;.
\end{equation}

\begin{figure}[ht!]
\plotone{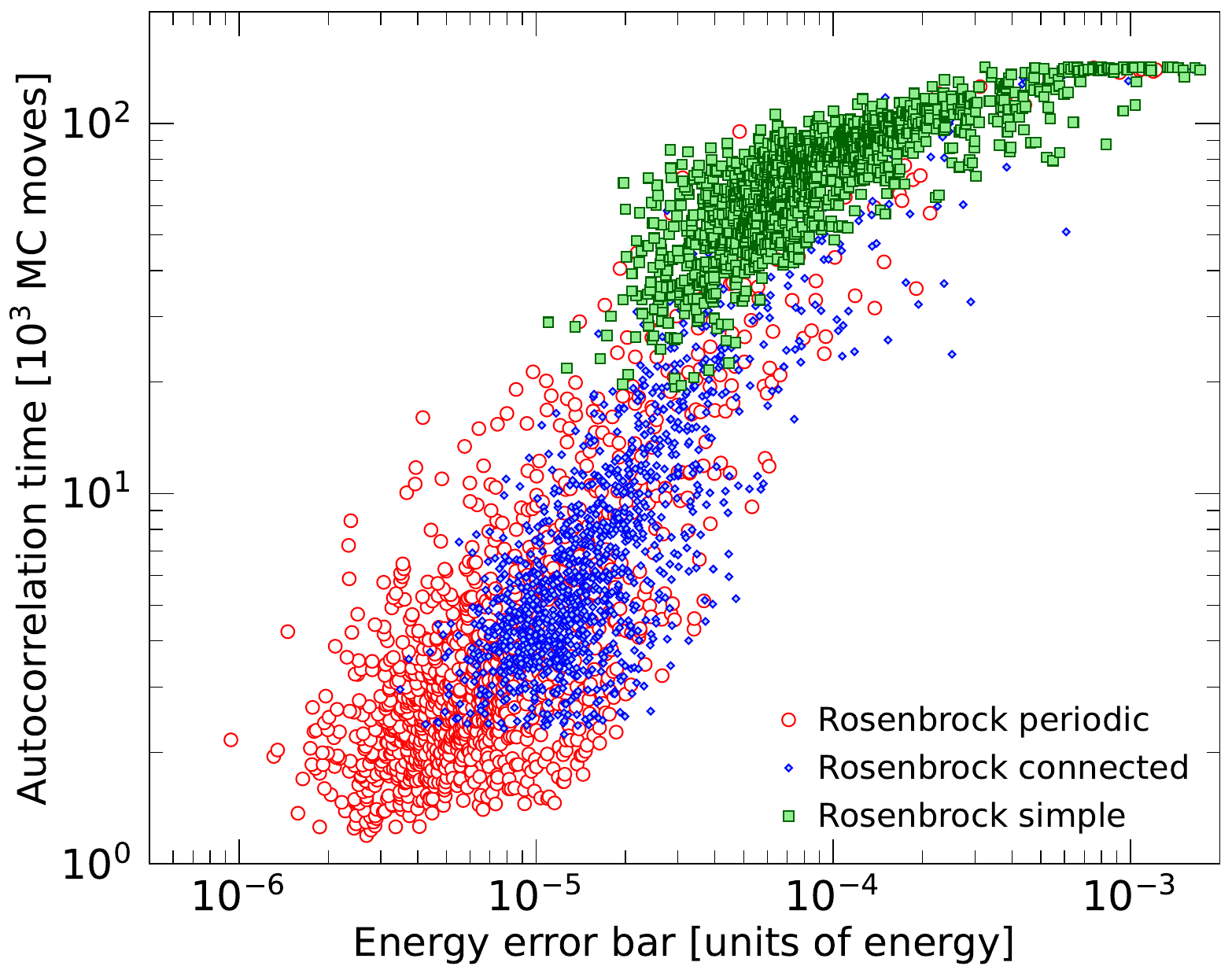}
\caption{Autocorrelation time and energy error bar that were computed with a variety of MC moves for the three types of Rosenbrock functions in ten dimensions. The plot reveals that the simple Rosenbrock function is most challenging to sample. For this reason, we focus on this function when we compare the performance of different MC methods in all following figures. \label{fig:rosen_potential_types}}
\end{figure}

In Fig.~\ref{fig:rosen_potential_types}, we compare error bars and
autocorrelation times that we computed with all the different MC moves
and parameters that we described in Sec.~\ref{sec:methods} for these
three types of Rosenbrock densities in $N=10$ dimensions. We find that
the simple Rosenbrock density to be the most challenging one to sample
because it leads to the largest error bars and the longest
autocorrelation times. Since we are interesting here in designing
algorithms that can solve challenging sampling problems, we focus all
following investigations on the simple Rosenbrock density and exclude
the two others from further consideration. The simple Rosenbrock
density was also studied \citet{HuijserGoodmanBrewer2022} who
demonstrated that the affine invariant method exhibits undesirable
properties in more than 50 dimensions.

Our code is open source. Examples and installation instruction are
available here: \citet{QMC_code2}. A simpler source code for our
quadratic Monte Carlo method is available here: \citet{QMC_code}.

\section{Results} \label{sec:results}

In this section we compare the efficiency of all types of MC moves for
different sampling parameters, one being the size the ensemble,
$N_W$. After \citet{Militzer_QMC_2023} demonstrated that setting $N_W$
larger than $3N+1$ leads to disproportionally long travel times, we
conduct simulations with $N_W=\{N+2,N+3,N+4,3N/2,2N+1,3N+1\}$ for all
method.

For the affine and the modified affine methods, we conducted
simulations with $a=\{1.2, 1.5, 2.0, 2.5\}$. For the affine simplex
method, we combined these four $a$ values with using
$N_G=\{3,4,5,6,10\}$ guide points to construct the simplex, which
brings the number of separate calculations to 20 for this method.

For the quadratic MC method with linear and Gaussian $t$ sampling, we
performed simulations for $a=\{0.1,0.3,0.5,1.0,1.2,1.5,2.0,3.0\}$. We
use the same set of $a$ values to conduct higher-order simulations
with $N_O=\{3,4,6,10\}$, again with both types of $t$
sampling. Finally we combined these eight $a$ value with
$N_G=\{3,4,5,6,10\}$ guide points to conduct simulations with the
quadratic simplex and with the walk method.

This typically led 1224 separate MC calculations for a particular
problem and a given dimension. For $N=2$ dimensional Rosenbrock
density, this only led to 508 separate MC simulations because in some
cases, the number of walkers was too small to allow one to chose the
specified number of guide points.

\subsection{Rosenbrock results}

\begin{figure}[ht!]
\plotone{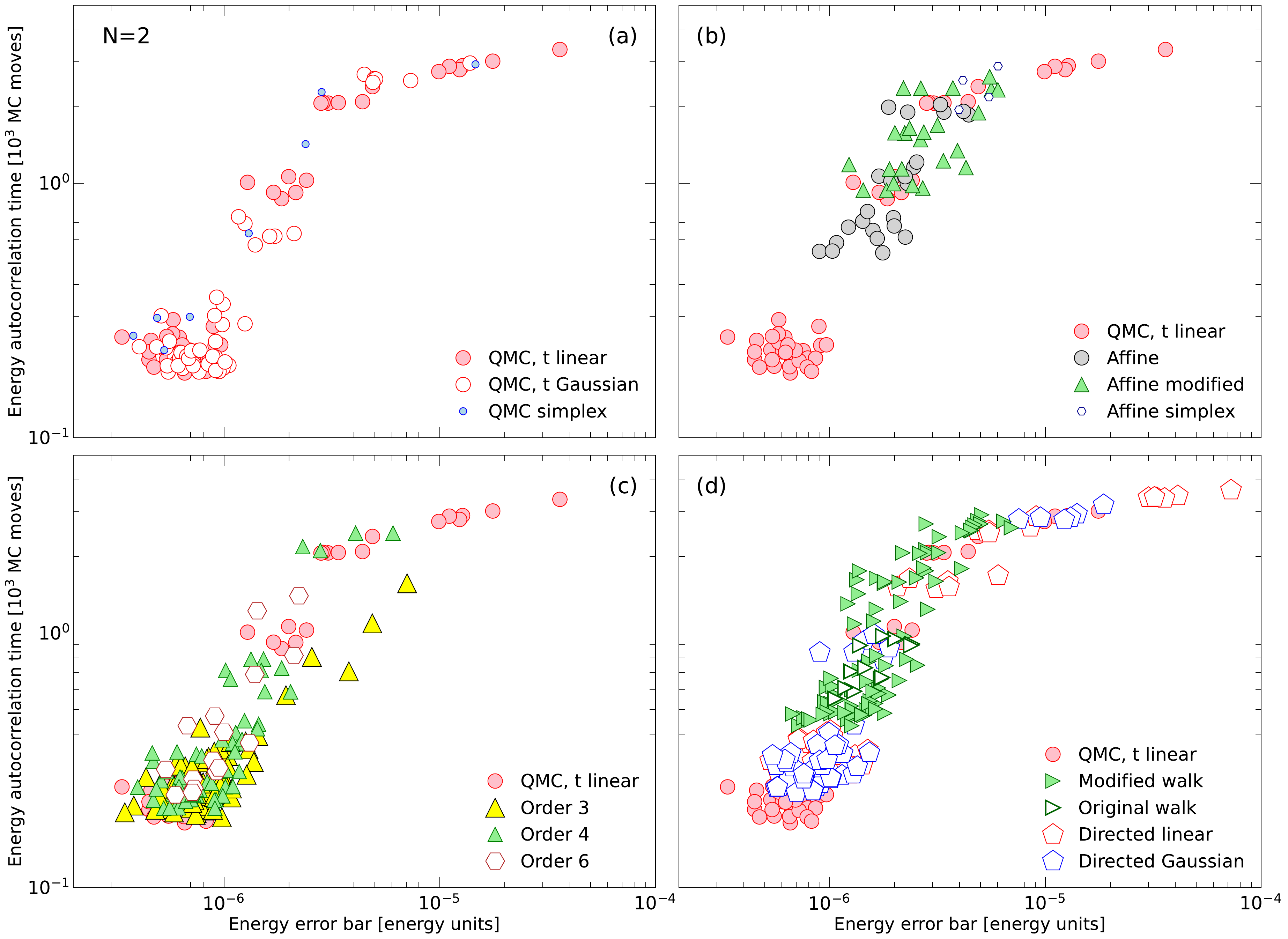}
\caption{Autocorrelation time and energy error bar that were computed
  for the Rosenbrock function in two dimensions. The four panels
  compare the performance of different MC methods and sampling
  parameters (see text). Results from our QMC method with linear
  sampling in $t$ space are repeated across all panels to simplify the
  comparison. To keep this and all following figures simple, we combine 
  results of the higher-order method that were obtained with 
  different $a$ values and with linear or Gaussian $t$ sampling 
  into a single dataset for a given order, $N_O$. 
  \label{fig:rosen_N=2}}
\end{figure}

\begin{figure}[ht!]
\plotone{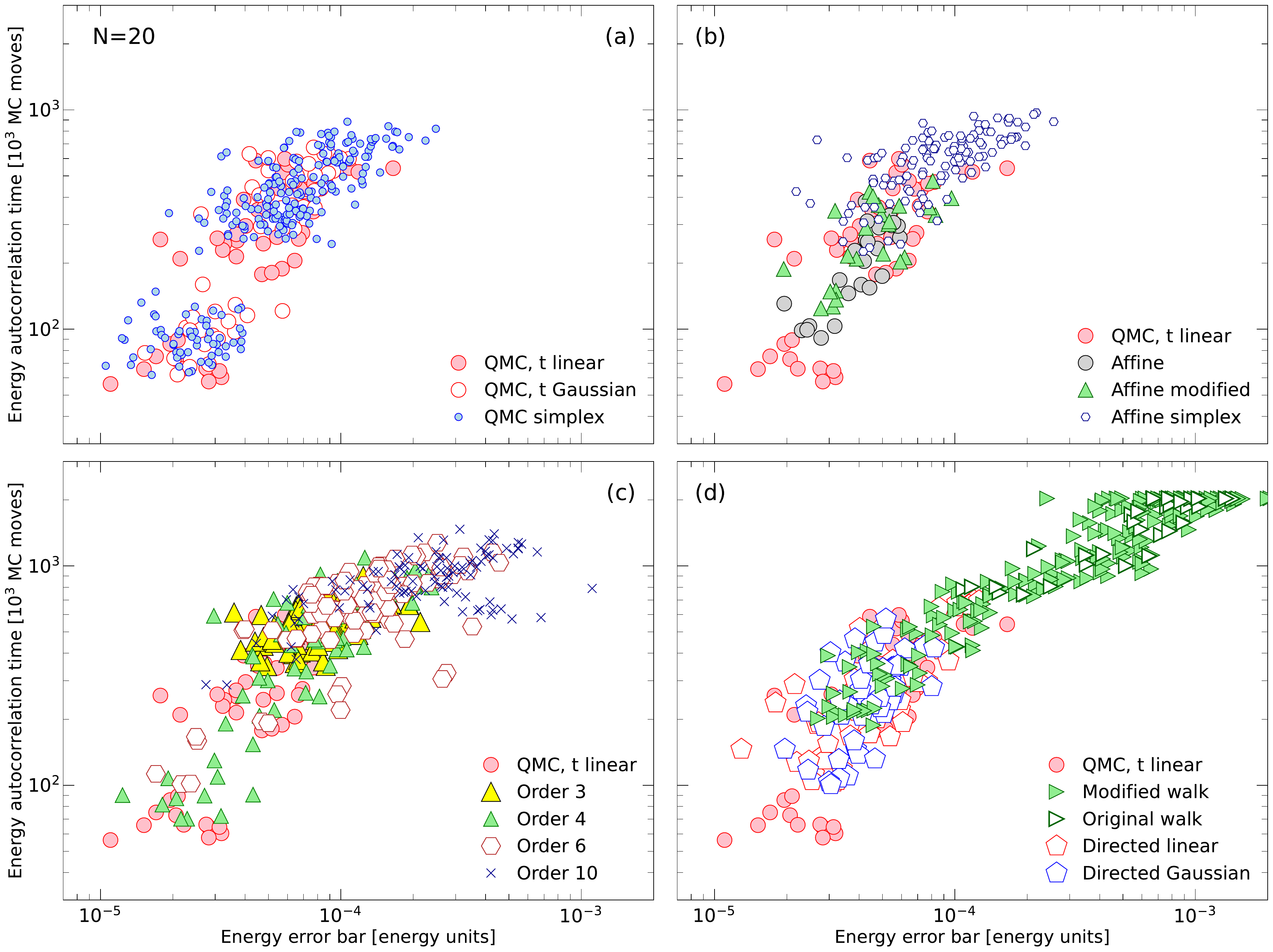}
\caption{Autocorrelation time and energy error bar that were computed for the simple Rosenbrock function in twenty dimensions. Like in Fig.~\ref{fig:rosen_N=2}, the four panels compare the performance of different MC methods. Results from our QMC method with linear sampling in $t$ space are again repeated across all panels for the comparison. \label{fig:rosen_N=20}}
\end{figure}

\begin{figure}[ht!]
\plotone{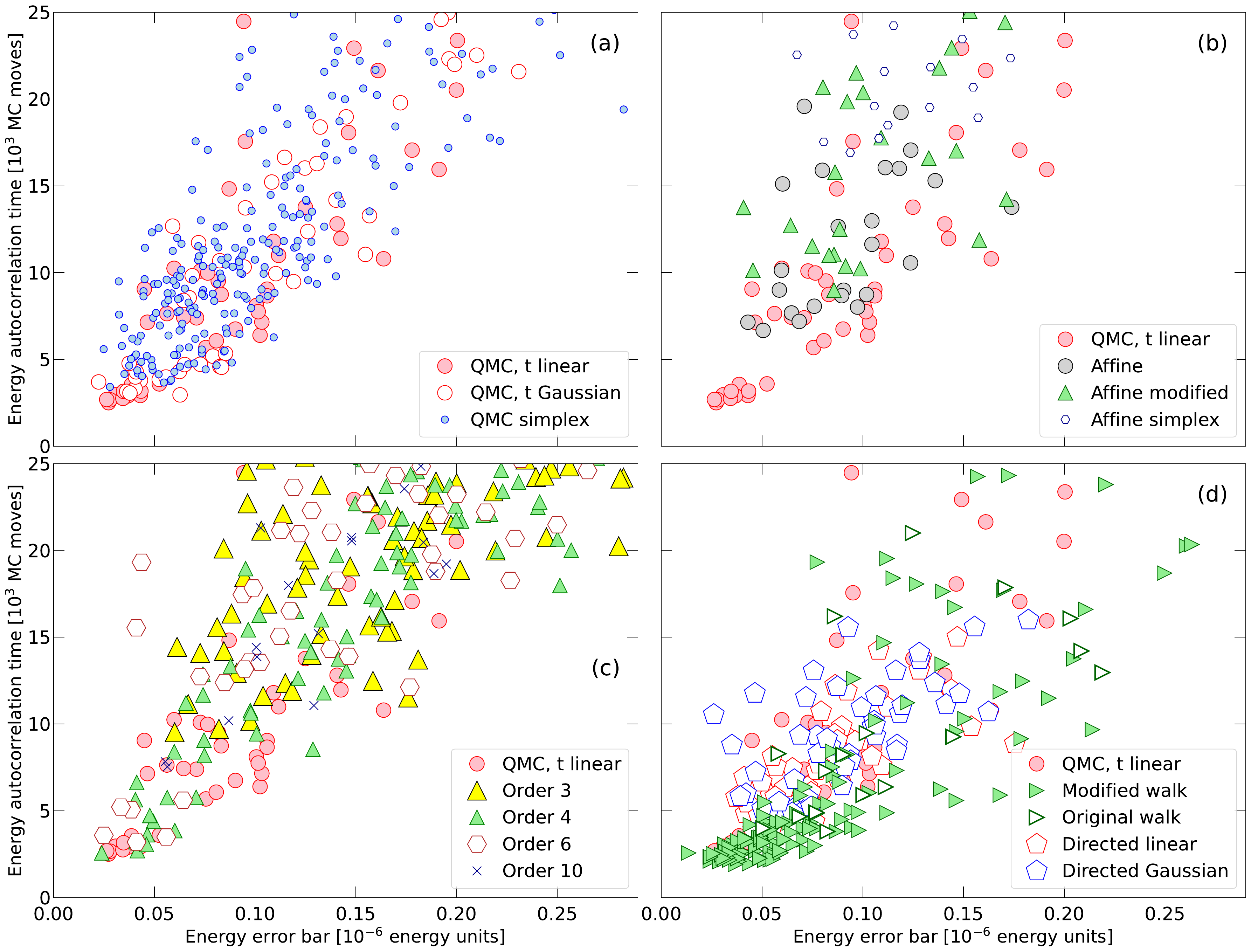}
\plotone{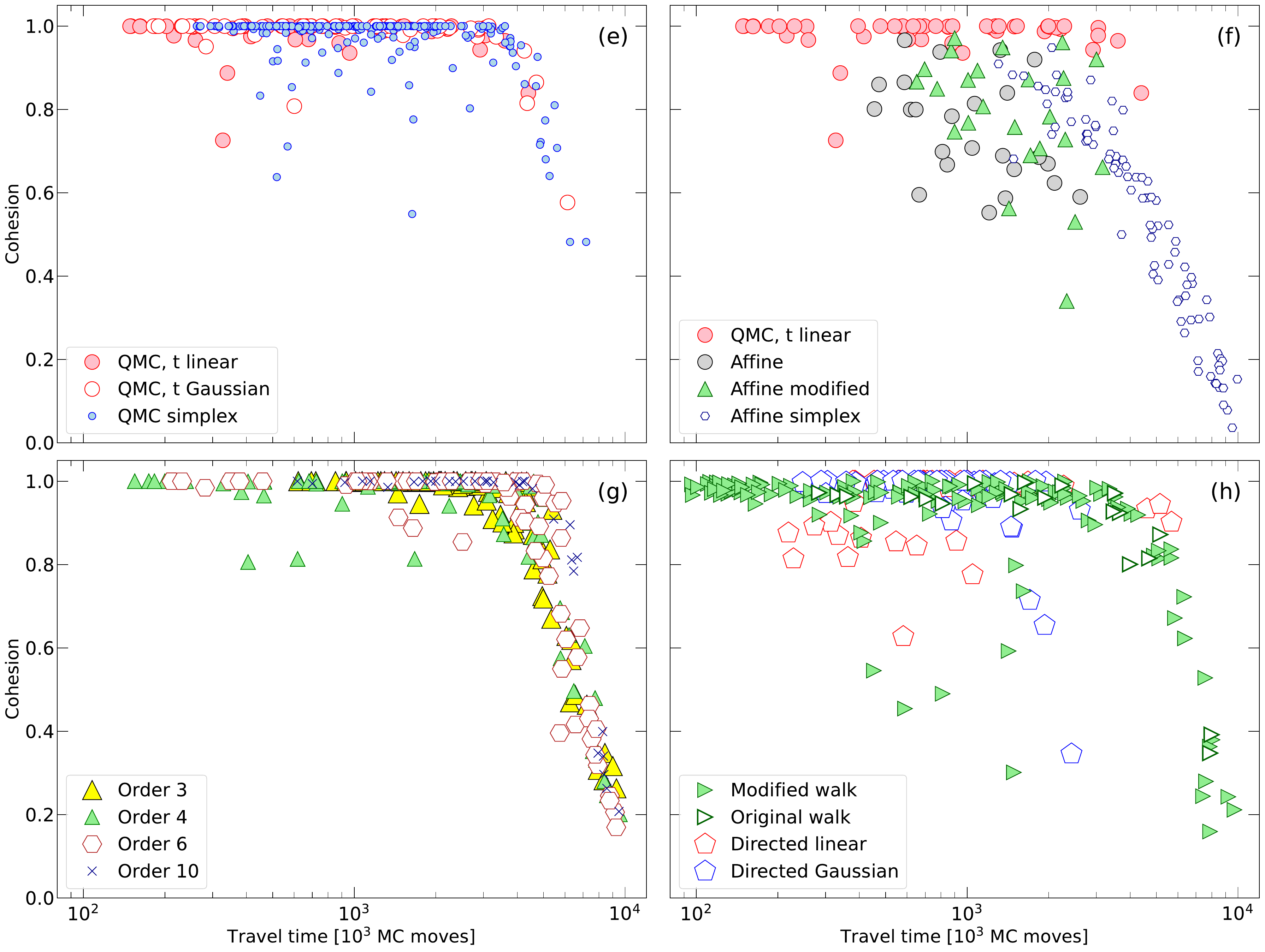}
\caption{Simulation results for the ring potential in N=24 dimensions. In the upper four panels, we compare the energy autocorrelation time and error bar from different sampling methods while we plot the cohesion and travel time results in lower panels. \label{fig:ring02}}
\end{figure}

In Figs.~\ref{fig:rosen_N=2} and \ref{fig:rosen_N=20}, we respectively
compare the results from different MC sampling methods for the
Rosenbrock function in $N=2$ and 20 dimensions. In
Fig.~\ref{fig:ring02}, we show similar results for the ring potential
in $N=24$. We also computed results for $N=12$ dimensions but we do
not provide a plot here because they are very similar, with one
exception: The order 3 results did not lag behind the QMC results in
$N=12$ case as they do for $N=24$ in Fig.~\ref{fig:ring02}. One sees
the same trend if one compares the $N=2$ and 20 results for the
Rosenbrock function.

For any application of MC methods, one wants the autocorrelation time
and computed averages as small as possible. We compute both for the
energy because this is the central quantity of many physical
system. The results from different sampling methods are compared in
panels (a)-(d) in Figs.~\ref{fig:rosen_N=2}, \ref{fig:rosen_N=20}, and
Fig.~\ref{fig:ring02}. For the ring potential, an efficient algorithm
should fulfill two additional criteria: It should travel efficiently
from unfavorable to favorable regions of the parameter space and
exhibit a high level of cohesion. Both are plotted in panels (e)-(h)
of Fig.~\ref{fig:ring02}.

The QMC method with linear and Gaussian $t$ sampling yield very good
results across all panels in Figs.~\ref{fig:rosen_N=2},
\ref{fig:rosen_N=20}, and \ref{fig:ring02} but there is also
quite a bit of variance among the predictions. So with the following
discussion, we identify favorable choice for the different sampling
parameters and compare for the two applications with goal of
identifying some reasonable default settings for future applications.

For Rosenbrock function in $N=2$ dimensions, the QMC method with
linear $t$ sampling works well if the scaling parameter $a$ is set
between 1.0 and 3.0 for any number of walkers between 3 and 7. If $a$
is chosen poorly between 0.1 and 0.5, the energy error bar and
autocorrelation time increase drastically, which is the trend we see
in Fig.~\ref{fig:rosen_N=2}a. Similarly, the QMC method with Gaussian
$t$ sampling works well for $a$ value between 0.5 and 3.0 and not so
well for $a=0.1$ and 0.3.

If we increase the number of dimensions from 2 to 20, the favorable
parameter ranges are slightly modified. The QMC method with linear $t$
sampling works well if $a$ is set only between 0.3 and 1.0 but not so
well for larger $a$ values, nor for $a=0.1$. The QMC method with
Gaussian $t$ sampling works well for $a$ values between 0.1 and 0.5
but not so well for larger $a$ values. As one increases the
dimensionality of the Rosenbrock density, sampling the search space
becomes more challenging~\citep{HuijserGoodmanBrewer2022} and one
needs to make smaller steps to still move efficiently.

When we compare the performance of the QMC method with linear $t$
sampling for the ring potential in $N=24$ dimension, we find the
method to yield small error bars, short autocorrelation and travel
times as well as a high level of cohesion of over 90\% if $a$ is set
between 0.3 and 0.5 for between 26 and 73 walkers. 

For $a=1.0$, the method works well for between 27 and 49 walkers.
If we switch to Gaussian $t$ sampling, $a$ values between 0.1 and 0.5
yield favorable results.
The QMC simplex method yields results that are very similar to that of
the original QMC method as Figs.~\ref{fig:rosen_N=2},
\ref{fig:rosen_N=20}, and \ref{fig:ring02} show consistently. This
also means that these test cases did not reveal any particular
advantage of this method.

We find that the affine invariant method does not perform as well as
the QMC method for all three test cases. It lags behind by factors
between 2 and 4 behind the best QMC results as panels (b) of
Figs.~\ref{fig:rosen_N=2}, \ref{fig:rosen_N=20}, and \ref{fig:ring02}
show. We obtained the best results for the ring potential by setting
$a=1.2$ and using between 28 and 73 walkers. For these simulations,
the cohesion level varied between 80 and 97\% (see
Fig.~\ref{fig:ring02}f). For the Rosenbrock density in 2 and 20
dimensions, we needed to set $a$ to 2.5 and 1.2 respectively to obtain
the best results. This confirms the trend that sampling the $N=20$
space efficiently requires smaller steps.

Our modification to the affine method did not yield any
improvements. In Figs.~\ref{fig:rosen_N=20} and \ref{fig:ring02}, it
slightly lags behind the original affine method while
Figs.~\ref{fig:rosen_N=2} shows the autocorrelation time for the 2d
Rosenbrock density is about twice as long. The affine simplex
performed yet a bit worth. It lagged behind the other affine methods
in term of the travel time and cohesion as Fig.~\ref{fig:ring02}f shows. 

\subsection{Ring potential}

We obtained the best results for the ring potential in $N=24$
dimensions in Fig.~\ref{fig:ring02} when we employed the QMC method
with linear $t$ sampling with $a=0.3\ldots0.5$ and Gaussian $t$
sampling with $a=0.1\ldots0.3$. Using between 27 and 73 walkers gave
consistently good results. In comparison, the quadratic simplex method
did not quite perform as well for any sampling parameters.

Affine and modified affine methods did slightly worse than the
quadratic simplex method. The autocorrelation time, error bars, and
travel times were not as good and the cohesion varied substantially
between 50\% and 95\% (see Fig.~\ref{fig:ring02}b and f). The best
choice for the $a$ value was 1.2. The affine simplex method performed
poorly on all counts.

In Fig.~\ref{fig:ring02}c and g, we compare results from
interpolations with different orders. Some simulations of fourth and
sixth order performed well if we set $a$ to 0.1 or 0.3. In comparison,
calculations of third and tenth order were not competitive. 

In Fig.~\ref{fig:ring02}d and h, we compare results obtained with the
directed quadratic method and with walk moves. With the directed
quadratic method, we obtained the best results by setting $a$ to 0.3
and 0.5 but they were not as good as those from original quadratic
method. On the other hand, the results from the walk method were
consistent better as long as we set the scaling parameter $a$ to 0.1
or 0.5, which underlines the importance this parameter that was
introduced only recently by \citet{Militzer_QMC_2023}.

\subsection{Relative Inverse Efficiency}

\begin{figure}[ht!]
\plotone{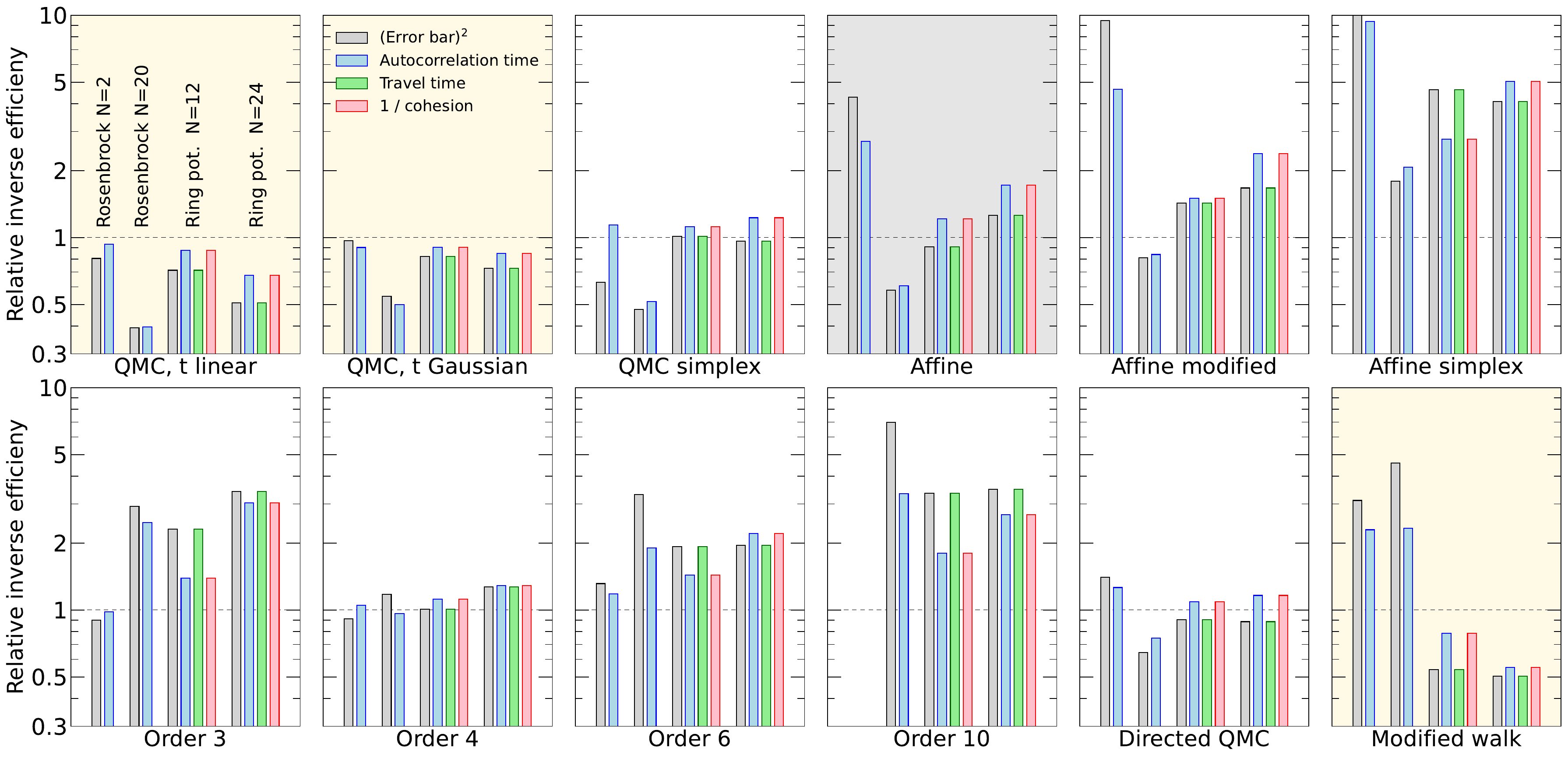}
\caption{The relative inverse efficiency (see text for definition) is shown for one MC method per panel. The four groups of vertical bars respectively show results from calculations for the simple Rosenbrock density in 2 and 20 as well as for the ring potential in 12 and 24 dimensions. The grey and blue bars represent the squared error bars and autocorrelation time of the computed energy. For the ring potential was also show the ensemble travel time arround the ring as well as the inverse of the ensemble cohesion so that small Y values imply a high efficiency in all cases. The QMC method with linear and Gaussian $t$ sampling demonstrated a high efficiency overall. The modified walk method performed better for the ring potential but did not yield competitive results for the Rosenbrock density. The yellow and grey backgrounds label methods that reported in \citet{Militzer_QMC_2023} and \citet{Goodman}, respectively. \label{fig:eff}}
\end{figure}

The preceeding analysis relied on a detailed comparision of results
spread across three figures. So here we introduce the {\em relative
  inverse efficiency} to greatly simplify the comparison of different
methods across various applications. Such an efficiency measure should
be able to handle the noise that we see in Figs.~\ref{fig:rosen_N=2},
\ref{fig:rosen_N=20}, and \ref{fig:ring02}. Furthermore, one cannot
expect that adjustable parameters like $a$ and $N_W$ have been
optimized to perfection. Instead it should be sufficient to have
results for good number of reasonable $a$ and $N_W$ values in the
dataset. Also a method should not be penalized very much if in
addition to results for reasonable parameters there are also some
findings for unreasonable values like $a=3$ in the dataset. Finally if
a method like the modified walk method has more adjustable parameters
and more calculations have been performed, this should not lead to an
advantage over other methods that have fewer adjustable
parameters. With these tree goals in mind, we define the relative
inverse efficiency to be the ratio, $A/B$. For given application
(e.g. the Rosenbrock density in 2 dimensions), we sort all
autocorrelation times obtained with all methods and all parameters by
magnitude and then compute the average time of the best $f=20\%$
results to obtain $B$. (This value would not change if a completely
useless method were to be added to the dataset.)  We derive $A$ by
averaging the best 20\% result derived with one particular method
only. Employing the fraction $f$ of the results rather than single
data points helps reduce the effect of the noise and it favors methods
to yield good results over a wider range of parameters, which is
helpful for real challenging application for which one may not be able
to perform many test calculations. The ratio, $A/B$, is thus a measure
how a particular method compares to the best of its peers. In
Fig.~\ref{fig:eff}, we plot a relative inverse efficiency for twelve
MC methods and four applications. We not only computed it for the
autocorrelation time but also for the squared error bar of the energy
and for the ring potential, also for the ring travel time and inverse
of the cohesion so that small values of all four measures imply a high
level of efficiency. (Setting $f=10\%$ did not yield any meaningful
change.)

Fig.~\ref{fig:eff} illustrates that the QMC method with linear and
Gaussian $t$ sampling has the highest efficiency overall, which is
consistently above average. The performance of the QMC simplex method
is also very good but it does not do quite as well for the ring
potential. Fig.~\ref{fig:eff} also illustrates that the original
affine method nor its modified or simplex forms can compete with the
QMC method. The efficieny is lower by factors between 2 and 10.

Furthermore Fig.~\ref{fig:eff} shows that increasing the interpolation
order from 2 (QMC) to 10 leads to a gradual decrease in efficiency but
this trend has one exceptions. The fourth-order method does notable
better than the third-order method, which only does well for the
Rosenbrock density in 2 dimensions.

The directed QMC method does very well but, for the four applications
presented here, it does not offer any improvement over the original
QMC method. Finally the modified walk method does markedly better than
the QMC moves for the ring potential according to all four efficiency
criteria but it does not sample the Rosenbrock density very well.

\section{Conclusions} \label{sec:conclusions} 

The work by \citet{Goodman} along with the practical implemenation by
\citet{emcee} made ensemble Monte Carlo calculations hugely
popular. Rather than relying on expert knowledge or on specific
properties of the fitness landscape to be sampled, the algorithm
harnesses information from the location of other walkers in the ensemble when
moves are proposed. \citet{Goodman} employed the {\em affine
  invariant} stretch moves from \citet{Christen} but also introduced
{\em walk} moves. \citet{Militzer_QMC_2023} recently modified the walk moves by
introducing a scaling factor, which made the sampling of challenging
fitness landscape more efficient by giving users some control over the
size of the moves. Furthermore \citet{Militzer_QMC_2023} introduced quadratic MC moves and
showed that it greatly improves the sampling of the Rosenbrock
function and a ring potential because the linear stretch moves of the
affine invariant method are not optimal for curved fitness landscapes.
This had brought the number of ensemble MC moves to three: {\em
  affine stretch} moves, {\em walk} moves, and {\em quadratic}
moves. Here are added five novel types of moves that we illustrated in
Fig.~\ref{fig:moves}. We introduced modified affine moves and affine
simplex moves. We added quadratic simplex moves but more importantly
generalized the quadratic moves to arbitrary interpolation
order. While we had success with the forth-order method, our analysis
also showed that increasing the interpolation order do not improve the
sampling efficiency of the Rosenbrock and ring potential functions. 
This conclusion is based on a relative inverse efficiency measure that
enables us to automatically compare sets of results from different MC
methods.

Besides requiring that a MC method leads to small error bars and short
autocorrelation times, we also require the ensemble to travel
efficiently from unfavorable to favorable region of the ring potential
fitness landscape while leaving no walkers behind, which we measure in
terms of {\em cohesion}. With this measure, we showed that the affine
method is more proning to leaving walkers behind than the quadratic
method.

Finally, we introduced the {\em directed quadratic moves}, which
differs from all other moves because we use the probability values in
addition to the location of the walkers, which is the sole piece of
information that the other moves rely on. The overall goal of this
manuscript is to broading the portfolio from three to eight types of
moves, that the popular ensemble MC methods rely on. We provide our
source code so that all moves can be inspected and their performance
be analyzed for other applications.

Not surprisingly, we find that there is no single MC method that
yields the best results in all situations. Furthermore, we confirmed
that all types of moves require at least an adjustment of a scaling
parameter, that controls the average step size, for the MC process to
run efficiently. There appears to be no perfect setting of sampling
parameters that is optimal for all problems.

If resources are very limited, we recommend comparing our quadratic
moves and modified walk moves for $a=\{0.5,1.0,1.5\}$ and then
refining the choice for $a$ in case various values yield very
different levels of efficiencies. Once a baseline case has been
established and resources are available, we recommended testing our
directed quadratic method as well as our fourth and sixth order
methods for parameters we discuss in section~\ref{sec:results}. In our
experience, it is sufficient to conduct such tests by choosing just
one value for the number of walkers between $2N+1$ and $3N+1$. Then
one should compare the resulting blocked error bars and auto
correlation time for quantities of high interest. Furthermore one
might want to check whether any walkers were stuck in unusual
parameter regions (cohesion).

\begin{acknowledgments}
  This work was supported by the Department of Energy-National Nuclear
  Security Administration (DE-NA0004147) via the Center for Matter at Extreme
  Conditions and by the Juno mission of the National
  Aeronautics and Space Administration.
\end{acknowledgments}


\end{document}